\documentclass[twocolumn,aps,prl,a4paper,nofootinbib]{revtex4}

\usepackage{graphicx}
\usepackage{fancyhdr}
\usepackage{amssymb}
\usepackage{pslatex}
\usepackage{multirow}

\begin{document}

\title {An Extended H\"uckel Study of the Electronic Properties of III-V Compounds and Their Alloys}

\author{Ingrid A. Ribeiro$^1$, Fabio J. Ribeiro$^2$ and A. S. Martins$^1$}

\affiliation{$^1$ Departmento de F\'\i sica, ICEx - Universidade Federal Fluminense}
\affiliation{$^2$ Departamento de Ci\^encias Exatas, EEIMVR - Universidade Federal Fluminense}

\begin{abstract}
In this work, we performed tight binding calculations of the electronic structure of III-V semiconductors compounds and their alloys based on the Extended H\"uckel Theory (EHT), where the H\"uckel parameters for the binary compounds were generated following a simulated annealing procedure. In particular, this article is focused on the dependency between band gap and the applied pressure and also the alloy composition.
\end{abstract}


\vskip -1.35cm

\maketitle

\thispagestyle{fancy}

\setcounter{page}{1}

\bigskip

\section{Introduction}
\label{intro}

III-V alloys are systems widely studied experimentally and theoretically, and some of these alloys display direct-indirect band gap crossover when the concentration $x$ of an element which belongs to III or V group is varied. In the theory front, these alloys are commonly studied by performing standard density functional theory (DFT) calculations based on both LDA and GGA approximation, although  sometimes they are not able to reproduce the critical concentration $x_c$ of the band gap crossover \cite{nicklas2010}. The DFT calculations demand the use of large supercells; moreover, in order to correctly represent the random dispersion of the elements in an alloy, the properties are calculated as an average of the values taken from an ensemble of configurations. This procedure leads to high computer demands despite the recent development of highly efficient \textit{ab-initio} DFT-based formalisms which scale linearly with the number of atoms.   

As an alternative procedure, it is common to approach the problem using semi-empirical tight-binding hamiltonians which employ linear combination of atomic orbitals (LCAO) for basis sets, with orbital energies and hopping parameters fitted to accurate band-structures as described in the seminal paper of Slater and Koster \cite{slater1954}. One example is the semi-empirical Extended H\"uckel Theory (EHT) method for electronic calculations in periodic and non-periodic systems, employed due to its simplicity and the chemical insight it can provide. The EHT presents the following advantages over the traditional orthogonal TB schemes: (\textit{i}) a considerable reduction in the number of parameters to be adjusted; (\textit{ii}) natural scaling laws for interaction among the atomic basis orbitals and, (\textit{iii}) an enhancement in transferability of parameters for different chemical environments. 

In this article is presented a reliable tight-binding (TB) study based on the Extended H\"uckel Formalism of selected III-V alloys electronic properties. Although the problem has been studied in different theoretical contexts like Orthogonal Tight-Binding (OTB) and \textit{ab-initio} formalisms, this is the first study within the Extended H\"uckel Theory (EHT). In order to establish advantages and drawbacks of the EHT, four III-V alloys were selected: $Al_{x}Ga_{1-x}As$, $GaAs_{1-x}P_{x}$, $Ga_{1-x}In_{x}As$, and $GaAs_{1-x}N_{x}$. These alloys present different levels of mismatch between the lattice parameters of the former compounds, being a good test to the ability of the EHT to take into account the effect of lattice distortion in the electronic gap values. Moreover, there are several previous theoretical and experimental studies on these alloys, allowing to establish the accuracy of EHT method.

The article is organized as follows. In the Section \ref{method}, a summary of the main ideas underlying the most common theoretical approaches to III-V alloys electronic structure calculations is presented, also followed by an introduction to the Extended H\"uckel Theory. Section \ref{comp} is dedicated to outline the Simulated annealing procedure for the H\"uckel parameters generation of the alloys's former binary compounds, together with a brief description of the computational details involving the electronic structure calculations of alloys in the TB frame. In the last sections the results are presented.

\section{Methodology}
\label{method}

Since the seminal work of Slater and Koster \cite{slater1954}, the tight-binding (TB) formalism has been a reliable tool for describing the electronic properties of crystalline solids. In the orthogonal formulation of the TB method, the crystalline states are described as linear combination of the atomic orbitals (LCAO), where the basis functions - the atomic orbitals for each specie - are not explicitly expressed in terms of known functions but they are used as a formal tool to construct all Hamiltonian matrix elements. Assuming that the atomic orbitals form an orthogonal basis set, the resulting orthogonal tight-binding (OTB) methods are quick and practical and its parameters, the orbital energies and hoppings, are adjusted in order to reproduce the band structure of the target material in a specific geometry. In this way, the TB parameters in the orthogonal formulation are usually not very transferable to different chemical environments and, in addition, the parameter values should be adjusted in the case of the system undergoes structural deformations. Tight-binding basis settings are commonly assumed to be both orthogonal and short ranged, while atomic wave functions are not, which implies that OTB basis sets do not resemble the eigenstates of an atomic Hamiltonian. Improvements for the efficiency in the OTB can be obtained by including hoppings beyond the first-neighbors and also adding more orbitals to the basis set. For systems under strain, where the atoms do not rest in the crystalline positions, a common way to approach the problem is correcting the hoppings using the power law Harrison scaling \cite{capaz1993}, but this procedure is successful only for small strain values.

On the other hand, first principle calculations yield accurate spectrum, but they are computationally demanding. In the first-principles techniques, the structural deformations are naturally taken into account in the total energy calculations by solving an one electron Schr\"odinger equation in a suitable self-consistent potential which approximates the electron-electron interaction. The Extended H\"uckel Theory (EHT) is a semi-empirical technique relying between the OTB and first-principle limits: the method works with explicit expressions for the basis orbitals. For a given geometry, the EHT basis functions are used to calculate a non-orthogonal overlap matrix $\mathbf{S}$, and this matrix and the fitted onsite orbital energies yield the corresponding off-diagonal hopping elements of the Hamiltonian. Within the standard H\"uckel prescription, structural changes are simply accounted by recalculating the overlap and hopping elements but leaving the basis sets and onsite elements unchanged\footnote{This approach yield good results only in the case for small values of structural deformation, that would justify do not recalculate the on-site energies}.

The most striking difference between EHT and OTB is that in EHT works with explicit atomiclike orbital (AO) basis functions, which are used to construct the matrix elements of $\mathbf{S}$ and $\mathbf{H}$. Compared to OTB, in the extended H\"uckel theory only the diagonal matrix elements of the Hamiltonian $H_{\mu\mu}$ (onsite energies) and the parameters specifying the basis functions are adjusted. In this article, the basis atomic orbitals are assumed to be spanned in terms of Slater-type orbitals (STO). Since the basis functions are known, the overlap matrix $\mathbf{S}$ is calculated explicitly and used to construct the off-diagonal matrix elements of the Hamiltonian hopping according to:
\begin{eqnarray}
 H_{\mu\mu} &=& E_{\mu\mu}  \nonumber \\
 H_{\mu\nu} &=& \frac{1}{2}K_{EHT}\left(H_{\mu\mu} +H_{\nu\nu} \right)S_{\mu\nu} \nonumber \\
 S_{\mu\nu} &=& \int{\phi_{\mu}^{\ast}\phi_{\nu}d^{3}\mathbf{r}},
\label{huckel}
\end{eqnarray}
where $K_{EHT}$ is an additional fitting parameter whose value is commonly set to 1.75 for molecules and 2.3 for solids \cite{jcerda2000}.

\section{Computational Details}
\label{comp}

In this work, the electronic structure of the III-V alloys and the corresponding former binary compounds were calculated within a non-orthogonal Slater-Koster scheme \cite{slater1954} using extended H\"uckel theory to derive the overlap and the Hamiltonian matrices $\mathbf{S}$ and $\mathbf{H}$, respectively. The analytical formulas of Michael Barnett \cite{barnett2003} were employed to calculate the $\sigma$, $\pi$ and $\delta$ components of the overlap among the basis orbitals. Within the non-orthogonal scheme, the band structure of a system was calculated by solving the generalized eigenvalue problem
\begin{equation}
\mathbf{H}(\mathbf{k})\Psi_{i}(\mathbf{k})=E_{i}(\mathbf{k})\mathbf{S}(\mathbf{k})\Psi_{i}(\mathbf{k}),
\label{gevp}
\end{equation}
where $\Psi_{i}(\mathbf{k})$ denotes the eigenvector of the $i$th band and $\mathbf{k}$ is the Bloch wave vector within the first Brillouin Zone. The overlap and Hamilton matrices, $S(\mathbf{k})$ and $H(\mathbf{k})$, were calculated by
\begin{eqnarray}
 H_{i,j}(\mathbf{k}) &=& \sum_{j',m'} e^{i\mathbf{k}\cdot\left(R_{i0}-R_{j'm'} \right)}H_{i0,j'm'}  \label{hhm} \\
 S_{i,j}(\mathbf{k}) &=& \sum_{j',m'} e^{i\mathbf{k}\cdot\left(R_{i0}-R_{j'm'} \right)}S_{i0,j'm'} \label{hsm},
\end{eqnarray}
where $i$ and $j$ label the atoms within the unit cell and $m'$ is the unit cell index. The sum in the equations \ref{hhm} and \ref{hsm} run over all atoms $j'$ in the unit cell $m'$ which are equivalent to atom $j$ in the reference unit cell $m=0$. The real-space matrix elements $H_{i0,j'm'}$ and $S_{i0,j'm'}$ between an atom $i$ in the reference unit cell and atom $j'$ in the cell $m'$ are calculated within of the extended H\"uckel prescription \ref{huckel}.

In order to describe an alloy, it is necessary to build a supercell hamiltonian, which implies the use of a large unit cell containing at least some hundreds of atoms. The considered supercells in this work were generated from a cubic unit cell of side $a$ with the atoms placed in sites of a zincblend lattice and they were built by setting $N_x$, $N_y$ and $N_z$, the number of unit (cubic) cells along the $x$, $y$ and $z$ directions, respectively, yielding a total number of atoms of $N=8N_{x}N_{y}N_{z}$. In this approach, all valence orbitals of the atoms belonging to the supercell enters in the basis set and the corresponding hamiltonian is not given by Eq. \ref{hhm}, built on a Bloch sums basis set, but a hamiltonian in the real space. Concerning the hoppings, they were restricted to sites with inter-atomic distances less than 9 $\textnormal{\AA}$ (cutoff radius).

Following Ref. \cite{jcerda2000}, a $spd$ set of valence orbitals $\{\Phi_{nlm}\}$ was built for each atomic specie, where the radial part of each orbital is spanned as a linear combination of two Slater-Type Orbitals (double zeta basis): 
\begin{equation}
\Phi_{nlm} = \sum_{i=1}^{2}c_{i}r^{n-1}e^{-\zeta_{i}r}Y_{lm}\left(\theta,\phi \right).
\label{stosum}
\end{equation}
For each atom type, there are three onsite energies ($E_s$, $E_p$ and $E_d$), the zetas and the first expansion coefficient $c_1$, providing a total of 12 parameters for atom type. The value of the $c_2$ coefficient is constrained in order to guarantee the normalization of the atomic orbital. In the Ref. \cite{jcerda2000}, there are TB parameterizations within EHT published for around 40 elements of the periodic table. In order to explore the effect of the parameterization in the alloy's gaps, in this work a new set of H\"uckel parameters for zincblend $GaAs$, $AlAs$, $InAs$, and $GaP$ and $GaN$ binary compounds is presented. The H\"uckel parameters were generated using a simulated annealing (SA) approach within the proposal of Vanderbilt \cite{vanderbilt1984}. In a few words, the SA consists in varying the H\"uckel parameters in successive Monte Carlo cycles with decreasing temperatures, aiming the reduction of an objective function $y$, defined as the root mean square (RMS) deviation of the H\"uckel bands $E_{i}^{H}(\mathbf{k})$ with respect to a target band structure $E_{i}^{T}(\mathbf{k})$:
\begin{equation}
y = \frac{1}{nb\times nk}\sqrt{\sum_{i=1}^{nb}\sum_{j=1}^{nk}\left[E_{i}^{H}(\mathbf{k}_{j}) - E_{i}^{T}(\mathbf{k}_{j}) \right]^2},
\label{objfunc}
\end{equation}
where $nb$ and $nk$ denote, respectively, the number of bands and k-points. The target band structured were calculated in the {\it ab-initio} density functional theory (DFT) formalism as implemented in the Abinit package \cite{gonze2009}, with plane wave cutoff energy of 40 Ha and Troullier-Martins pseudopotentials~\cite{martins1991}. These calculations were carried out by using the local density approximation (LDA) as parameterization for the exchange-correlation potential and the bands were generated along the $\Gamma-X-L-\Gamma$ lines. The band gaps of the DFT target bands were posteriorly corrected by shifting their conduction bands by the difference between the correct band gap and the DFT one. In the minimization procedure, all valence bands and the two first conduction bands were included in Eq. \ref{objfunc} and an acceptable set was generated when $y \approx 0.1$ eV.

\section{Results and Discussion}
\label{results}

\subsection {Results for the Binary Former Compounds}

The tables \ref{table:tab1} and \ref{table:tab2} present the H\"uckel parameters for the binary compounds that form the studied alloys. All parameters were obtained by following the SA approach exposed in the previous section and yielded final H\"uckel bands with RMS$\approx 0.1$ eV with respect to the DFT target bands. For all compounds, $K_{EHT}=2.3$ and the Fermi Level was fixed in $-13$ eV. For the parameterization procedure, it was assumed the transferability of the atomic orbitals (AO) parameters, constraining the $\zeta$ and the  $c$ values of a given atomic specie to be the same independent of the compound. Although all AO were spanned in a double-$\zeta$ basis, there were cases that the AO was well described with just one Slater orbital. In these cases, the values of $\zeta_2$ was set to a large value, $\zeta_2 = 25$, and the value of the $c_2$ coefficient, which ensures the AO normalization, was given by $c_2=\sqrt{1 - c_1^{2}}$ which means no overlap among the Slater basis orbital and their neighbors. 

\begin{table}
\scriptsize
\caption{\scriptsize Optimized parameters of the atomic orbitals (AO) basis set calculated by SA. Although all AO's are of the double-$\zeta$ Slater type, the values of the $c_2$ coefficient is not included whenever $\zeta_{2} = 25$ (see text). The $K$ value in Eq. \ref{huckel} was set to 2.3 and the Fermi Level was fixed to -13 eV.}
\begin{tabular}{|l|c|c|c|c|c|}
\hline\hline
   & AO & $\zeta_{1}$ & $c_1$ & $\zeta_{2}$ & $c_2$ \\ \hline
\multirow{5}{*}{N}  & $2s$ & 2.489 & 0.994 &  & \\ & $2p$ & 1.890 & 0.889 & 3.659 & 0.458 \\ & $3d$ & 0.869 & 0.583 &  &  \\ \hline
\multirow{5}{*}{Al} & $3s$ & 1.493 & 0.644 &  & \\ & $3p$ & 1.278 & 0.690 & 4.538 & 0.724 \\ & $3d$ & 1.007 & 0.658 & 5.044 & 0.753 \\ \hline
\multirow{5}{*}{P}  & $3s$ & 2.509 & 0.987 &  & \\ & $3p$ & 1.631 & 0.437 & 2.783 & 0.899 \\ & $3d$ & 0.794 & 0.789 &  &  \\ \hline
\multirow{5}{*}{Ga} & $4s$ & 1.846 & 0.564 &  & \\ & $4p$ & 1.605 & 0.678 & 5.549 & 0.735 \\ & $4d$ & 1.287 & 0.695 &  &  \\ \hline
\multirow{5}{*}{As} & $4s$ & 2.649 & 0.765 &  & \\ & $4p$ & 2.047 & 0.783 & 7.539 & 0.622 \\ & $4d$ & 0.931 & 0.688 & 3.654 & 0.725 \\ \hline
\multirow{5}{*}{In} & $5s$ & 2.096 & 0.569 &  & \\ & $5p$ & 2.142 & 0.956 & 8.102 & 0.294 \\ & $5d$ & 1.051 & 0.662 & & \\
 \hline\hline
\end{tabular}
\label{table:tab1}
\normalfont
\end{table}

\begin{figure}[ht]
\begin{center}
\includegraphics[height=7.0cm]{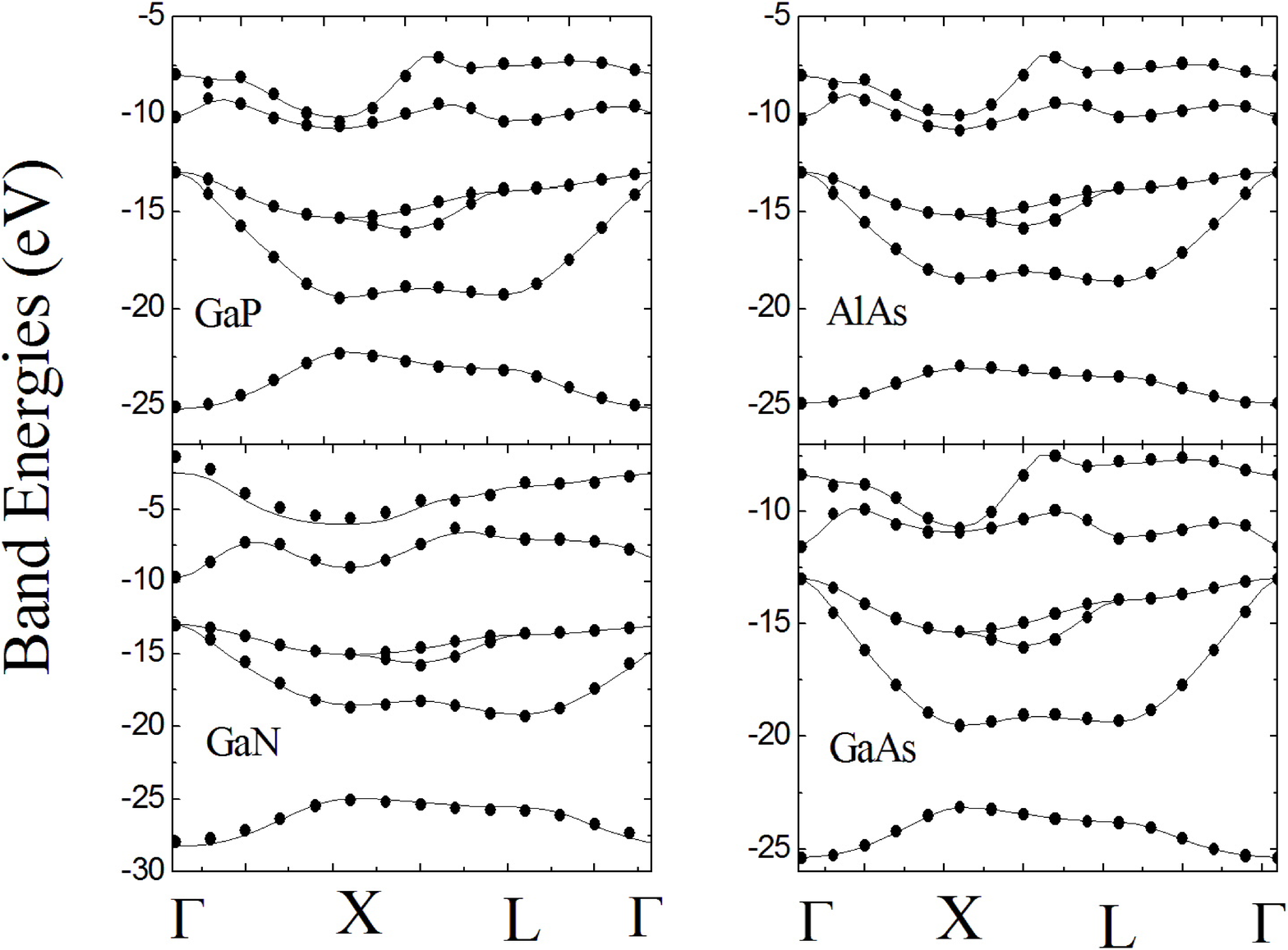}
\caption{Band structure of the binary compounds. The dots correspond to the target DFT LDA calculations, with the conduction bands rigidly shifted in order to correct the value of the gap. The solid lines corresponds to the calculated H\"uckel band structure. \label{fig1}}
\end{center}
\end{figure}

Concerning the on-site energies, for a given structure their values are dependent of the chemical environment as it can be observed in the table \ref{table:tab2}. The resulting band structures of four parameterized compounds are shown in Fig. \ref{fig1}, with the lines being the H\"uckel band structures calculated with the parameters from Tables \ref{table:tab1} and \ref{table:tab2} and the dots corresponding the target (gap corrected) are the LDA band structures. Notice the excellent agreement between the approaches.

\begin{table}
\tiny
\caption{\scriptsize Optimized on-site energies. The values of the band gaps and the lattice constants were taken from the reference \cite{vurgaftman2001}. Notice the excelent agreement among the experimental values for the gap compared with the values calculated using the optimized H\"uckel parameters.}
\begin{tabular}{|l|c|c|c|c|c|c|c|}
\hline\hline
  Compound & $a({\textnormal\AA})$ & $E_{g}^{Exp}$(eV) & $E_{g}^{Huc}$(eV) & Element & $E_s$ & $E_p$ & $E_d$ \\ \hline
  AlAs     &        5.66           &      2.23         &      2.24         &    Al   &  $-15.138$  &  $-9.813$  &  $-5.015$   \\
           &                       &                   &                   &    As   &  $-20.994$  &  $-12.535$ &  $-4.385$   \\
  GaAs     &        5.65           &      1.43         &      1.47         &    Ga   &  $-16.914$  &  $-9.767$  &  $-4.191$   \\
           &                       &                   &                   &    As   &  $-21.171$  &  $-12.629$ &  $-5.263$   \\
  InAs     &        6.04           &      0.42         &      0.43         &    In   &  $-16.727$  &  $-9.317$  &  $-4.819$   \\
           &                       &                   &                   &    As   &  $-21.575$  &  $-12.357$ &  $-5.429$   \\
   GaP     &        5.45           &      2.35         &      2.31         &    Ga   &  $-16.833$  &  $-9.683$  &  $-3.737$   \\
           &                       &                   &                   &    P    &  $-19.952$  &  $-12.526$ &  $-5.367$   \\
   GaN     &        4.54           &      3.30         &      3.30         &    Ga   &  $-15.682$  &  $-8.982$  &  $-3.662$   \\
           &                       &                   &                   &    N    &  $-22.907$  &  $-13.130$ &  $-3.305$   \\
 \hline\hline
\end{tabular}
\label{table:tab2}
\normalfont
\end{table}

As previously mentioned, in this article was assumed the transferability of the atomic basis orbitals. Thus, only the on-site energies were adjusted in order to take into account the following situations: (\textit{i}) the changing of chemical environment around the atoms in the alloys; and (\textit{ii}) the changing of the inter-atomic distances due the hydrostatic pressure application or to account the lattice distortions in the alloys. However, in a first approximation, the adjustment of the on-site energies for the second situation is not necessary, because the hopping parameters calculated by \ref{huckel} are proportional to the overlap $S_{\mu\nu}$ between the orbitals. One way to check this statement is by calculating the pressure coefficient $a_{P}^{\alpha}$ of an inter-band transition $\alpha$, defined as
\begin{equation}
a_{P}^{\alpha} = \frac{dE_{\alpha}}{dP}.
\label{defpot1}
\end{equation}
This coefficient is related to the volume ($V$) deformation potential
\begin{equation}
a_{V}^{\alpha} = \frac{dE_{\alpha}}{d \textnormal{ln}(V)},
\label{defpot2}
\end{equation}
with coefficient of proportionality given by the bulk modulus $B$,
\begin{equation}
a_{P}^{\alpha} = -\left( \frac{1}{B}\right)a_{V}^{\alpha}.
\label{defpot3}
\end{equation}
Changing the volume $V$ in the interval $-0.05 \le \textnormal{ln}(V_0/V) \le 0.20$, the values of the deformation potential were calculated from the slope of the $E_{gap}^{\alpha} \times \textnormal{ln}(V_0/V)$ curve for each compound.  In the Table \ref{table:tab3}, the results for the $\alpha=\Gamma_{8v}\rightarrow\Gamma_{6c}$ transition were summarized: apart the GaP compound, an excellent agreement between the calculated and the experimental values can be observed. It is important to emphasize the following point: in the calculations any additional parameter was not employed, just the hoppings were re-scaled by re-calculating the overlap matrix for each deformation. For the GaP, a better agreement can be found by re-calculating the H\"uckel parameters.

\begin{table}
\scriptsize
\caption{\scriptsize Deformations potentials for the $\Gamma-\Gamma$ transition. The physical unities of the bulk modulus and the $a_P$ are, respectively, $kBar$ and $meV/kBar$. The experimental values were taken from the \cite{zunger1999}}
\begin{tabular}{|l|c|c|c|c|}
\hline\hline
  Compound & $a_{V}^{\Gamma-\Gamma}$ & $B_{exp}$ & $a_{P}^{\Gamma-\Gamma}$  & $a_{P}^{\Gamma-\Gamma}$ (exp) \\ \hline
  GaAs     &   -8.17   &   756   &   10.81    &  8.5-12.6  \\
  AlAs     &   -8.57   &   781   &   10.97    &   10.2     \\
  InAs     &   -6.95   &   579   &   12.00    &   9.6-11.4 \\
   GaP     &  -11.12   &   882   &   12.61    &   9.7      \\
   GaN     &   -8.72   &  2054   &    4.24    &   4.0      \\
 \hline\hline
\end{tabular}
\label{table:tab3}
\normalfont
\end{table}

\subsection {Results for the III-V alloys}

Regarding the alloy calculations, it was employed supercells with $N_{x}=N_{y}=N_{z}=4$, where $N_{x}$, $N_{y}$ and $N_{z}$ are, respectively, the number of cubic cells of side $a$ along the $x$, $y$ and $z$ directions, resulting a number $N=8N_{x}\times N_{y}\times N_{z}=512$ of atoms. Two kinds of III-V alloys were considered: $A_{x}B_{1-x}C$ ($AlGaAs$ and $InGaAs$) and $AB_{1-x}C_{x}$ ($GaAsP$ and $GaAsN$). The alloy lattice parameter was calculated by using the Vegard Law, 
\begin{equation}
a = x\cdot a_{AC} + (1-x)\cdot a_{BC},
\label{vegard}
\end{equation}
employing a $A_{x}B_{1-x}C$ alloy as a reference. 

Concerning the atomic parameters from the tables \ref{table:tab1} and \ref{table:tab2}, it is important to discuss the employed model for the on-site energies in the alloy. Taking the $Al_{x}Ga_{1-x}As$ alloy as reference, the on-site energies for the common element in the former binary compounds, As, were calculated as a weighted average of the values for the former compounds, using an expression similar to the Vegard Law for the lattice parameter, Eq. \ref{vegard}. For the Al and Ga atoms in an alloy, the values of their on-site energies were set equal to the same average, which implies that the atoms partially lost their individuality and become a "pseudo" atom with orbital energies between the former values, in a similar fashion to the Virtual Crystal Approximation (VCA)\cite{vca}. However, only the on-site energies were varied: the zetas and the expansion coefficients were kept at their original values given in the Table \ref{table:tab1}.

Random alloys, as were considered in this article, lack formal translational symmetry and thus $\bf k$ is not a good quantum number, leading an inadequacy of the language of band-structure dispersion $E(\bf{k})$ to describe the energy states of the alloys. Nevertheless, several theoretical approaches have been proposed in the literature\cite{vca,boykin2007,boykin1,boykin2,popescu1,popescu2}, intending to restore the relation between the energy and $\bf k$. The Virtual Crystal Approximation\cite{vca} (VCA) was one of the first approximation employed in the theoretical study of $A_{x}B_{1-x}C$ semiconductor alloys into the TB frame, where the A and B atoms are replaced by a fictitious atom, whose TB parameters are calculated as weighted averages of the AC and BC binary parameter values. More realistic approaches, where the atom identity is preserved, are based on the spectral decomposition of the alloys eigenstates\cite{popescu1,popescu2} or the unfolding of the supercell Brillouin zone \cite{boykin2007,boykin1,boykin2}: the former approach employing plane waves as basis set and the second localized orbitals.

In this work, we proceeded a direct checking of the lowest unoccupied states of the calculated alloy spectrum. For the former binary compounds, corresponding to $x=0$ or $x=1$ alloy limits, the bottom of the conduction band can be non degenerate ($\Gamma$) or three-fold degenerate ($X$). On the other hand, for concentration values in the interval $0<x<1$, the degeneracy is partially lifted due to the spatial disorder, even though in this case is possible to identify the $\Gamma$, $X$ and $L$ character of the states: their energies lies between the $x=0$ and $x=1$ energy limit values.  

\subsubsection{$AlGaAs$ and $GaAsP$}

\begin{figure}[ht]
\begin{center}
\includegraphics[width=6cm,height=8cm]{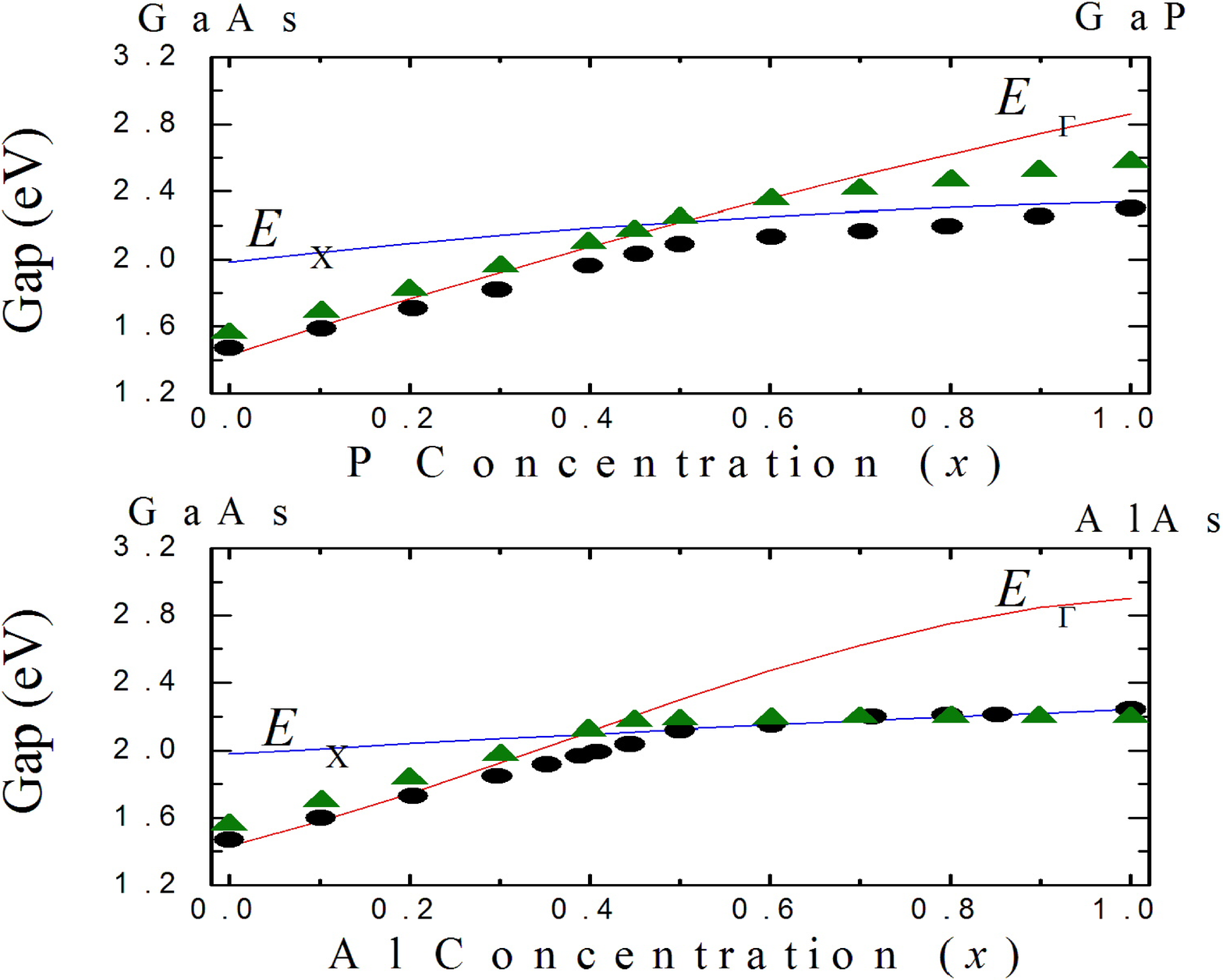}
\caption{Dependency of the gap energy with $x$ for the  $AlGaAs$ and $AlGaP$ alloys. The solid lines correspond to experimental values \cite{vurgaftman2001} and the circles and triangles corresponds, respectively, to the calculated H\"uckel's values for the J. Cerd\'a and the present parameterization (tables \ref{table:tab1} and \ref{table:tab2}).  \label{fig2}}
\end{center}
\end{figure}

Figure \ref{fig2} shows the results for the gap variation in $X$ and $\Gamma$ with respect to the concentration $x$ for the $AlGaAs$ and $AlGaP$ alloys. The points denote the theoretical results with the circles corresponding to the calculations using the presented H\"uckel parameters, tables \ref{table:tab1} and \ref{table:tab2}, and the triangles were calculated using the J. Cerd\'a parameters\cite{jcerda2000}. The solid lines correspond to fits of the experimental data \cite{vurgaftman2001} for the gaps in $X$ and $\Gamma$ and, as it can be observed in Fig. \ref{fig2}, the results are in good agreement with the experimental values, but they are sensible to the H\"uckel parameters. Both parameterizations yield good values for crossover concentration $x_c$: for $GaAsP$, $x_c$ lies between 0.45 and 0.50 and, for $AlGaAs$, the proposed parameterization yields $x_c \approx 0.5$ and J. Cerd\'a's parameterization gives $x_c \approx 0.4$, being the experimental value closer to the lated $0.38$. Concerning the $GaAsP$ alloy, the former compounds have a mismatch of $3.5\%$, and the inclusion of the atomic positions relaxation can improve the agreement between the calculated and the experimental results.

\subsubsection{$InGaAs$ and $GaAsN$}

$InGaAs$ and $GaAsN$ alloys are characterized by the large mismatch between the lattice parameters of the former compounds, $7\%$ and $20\%$, respectively. The $InGaAs$ alloy, in contrast with $AlGaAs$ and $GaAsP$, remains a direct-gap material over its entire composition range. With respect to  $GaAsN$, it is argued that the large miscibility gap between the former compounds become difficult to prepare alloys with large $N$ fractions, therefore is expected a phase separation in GaN-rich alloys \cite{vurgaftman2001}, in order that can be done only comparisons between theoretical previsions using different formalisms for this alloys. Moreover, it has been known that small quantities of nitrogen form deep-level impurities in $GaAs$ and $GaP$ and, in these cases, a different behavior of the band gap variation with the nitrogen concentration $x$ can be expected.

\begin{figure}[ht]
\begin{center}
\includegraphics[width=5.5cm,height=7.5cm]{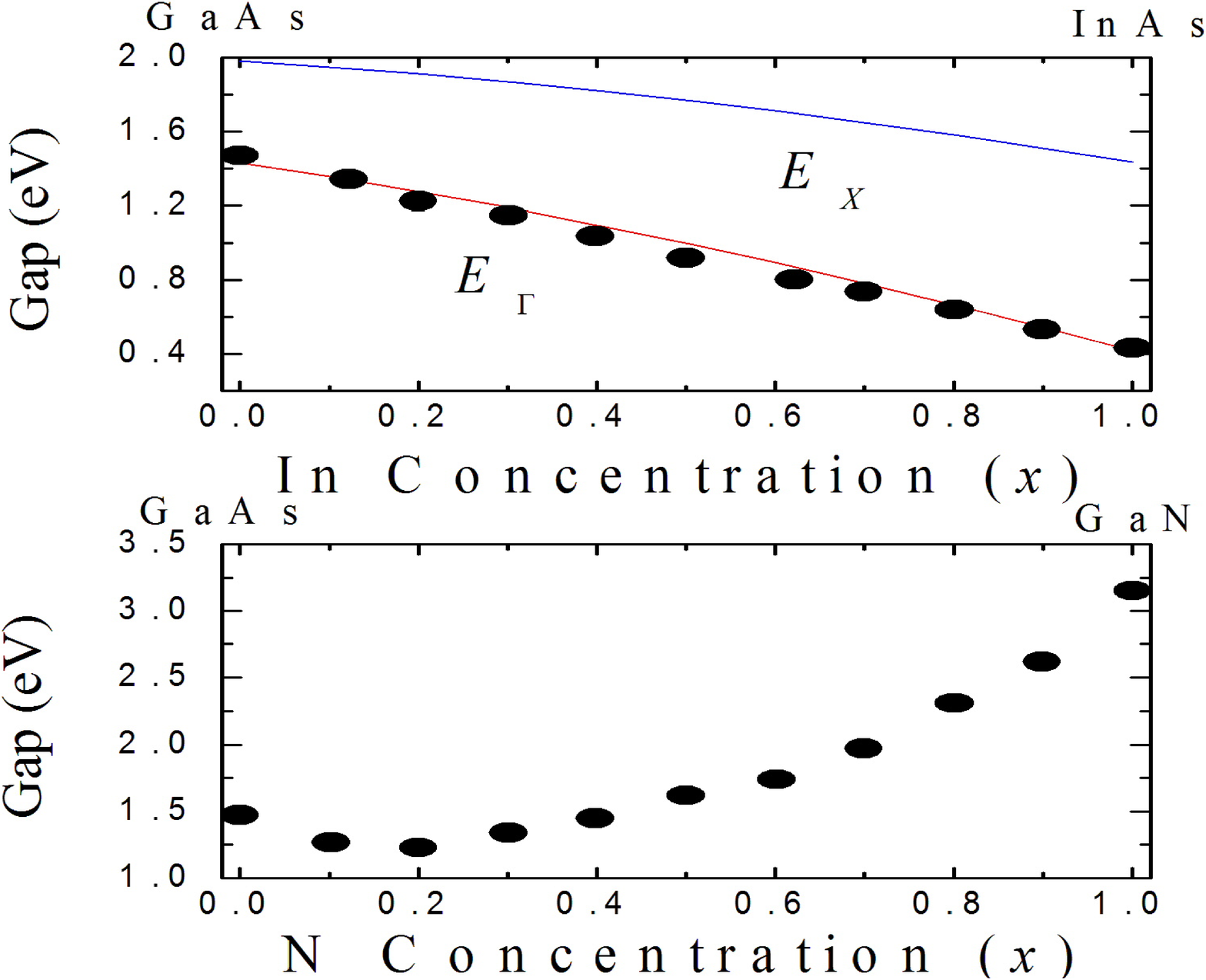}
\caption{Dependency of the gap energy with $x$ for the  $InGaAs$ and $GaAsN$ alloys.  \label{fig3}}
\end{center}
\end{figure}

Figure \ref{fig3} presents the results of the H\"uckel calculations: in the case of the $InGaAs$ alloy, it can be observed that the points remains close to the experimental values (line). In the case of the $GaAsN$, the H\"uckel calculations were able to reproduce some trends of the results calculated by Bellaiche {\textit{et al}}\cite{bellaiche1996}, in particular the band gap decreasing with $x$ for small values of the concentration, meaning a signature of deep levels in the band gap. Again, the inclusion of atomic relaxations may improve the agreement between the results.

\section{Conclusions}

In this article we presented a study of the dependency of the gap energy with the concentration $x$ for some semiconductor alloys in the context of the extended H\"uckel formalism. Although there were published H\"uckel parameters for among five binary III-V compounds \cite{jcerda2000}, this article presents new sets of parameters where their values were obtained by employing a simulated annealing procedure and their ``confidence" was tested by calculating the deformation potential in $\Gamma$ and the pressure coefficient, and as a result, a good agreement with the experimental values without need any additional parameter was shown. Concerning the alloys, a good agreement with the experimental values was achieved for the $AlGaAs$, $GaAsP$ and $InGaAs$ alloys, but the results comparison for the H\"uckel calculations by using two distinct parameterizations yielded slightly different theoretical curves for $AlGaAs$ and $GaAsP$. Regarding to the $InGaAs$ and $GaAsN$ alloys, the H\"uckel results reproduced well the experimental curve and, for $GaAsN$, the H\"uckel calculations were able to present the correct tendency of the gap variation with $x$ compared to the DFT study of Bellaiche {\textit{et. al.}}.

{\bf Acknowledgments}

The authors are grateful for the financial support of the Brazilian agencies Conselho Nacional de Desenvolvimento Cient\'{\i}fico e Tecnol\'ogico, CNPq, and of Funda\c c\~ao de Amparo \`a Pesquisa do Estado do Rio de Janeiro (FAPERJ). A. S. Martins would like specially to acknowledge Jorge I. Cerd\'a for all support concerning the implementation of the Extended H\"uckel Method and for all discussions about the H\"uckel parameters generation. This article is dedicated to the memory of the Professor Michael Barnett, who gave to the authors a precious help concerning the overlap integrals.

\end{document}